\documentclass{article}
\usepackage{spconf,amsmath,graphicx}

\usepackage{subfig}
\usepackage[table,xcdraw,dvipsnames]{xcolor}
\usepackage{multirow}
\usepackage{setspace}
\usepackage{rotating}
\newcommand\tabrotate[1]{\begin{turn}{90}\rlap{#1}\end{turn}}
\usepackage{hhline}

\usepackage{tikz}
\usepackage{pgfplots, pgfplotstable}
\pgfplotsset{compat=1.5.1}
\usetikzlibrary{shapes, arrows, calc, patterns, matrix, positioning}
\usepackage{ifthen}
\usepgfplotslibrary{groupplots}

\newcommand\numberthis{\addtocounter{equation}{1}\tag{\theequation}}

\def\x{{\mathbf x}}

\title{Content Adaptive Wavelet Lifting for Scalable Lossless Video Coding}
\name{Daniela Lanz, Christian Herbert, and Andr\'{e} Kaup}
\address{Multimedia Communications and Signal Processing\\
	Friedrich-Alexander-University Erlangen-N\"{u}rnberg (FAU)\\ 
	Cauerstr. 7, 91058 Erlangen, Germany\\
	Email: \{Daniela.Lanz,Christian.Herbert,Andre.Kaup\}@FAU.de}
\begin{document}
\ninept
\maketitle
\begin{abstract}
Scalable lossless video coding is an important aspect for many professional applications. 
Wavelet-based video coding decomposes an input sequence into a lowpass and a highpass subband by filtering along the temporal axis. The lowpass subband can be used for previewing purposes, while the highpass subband provides the residual content for lossless reconstruction of the original sequence.
The recursive application of the wavelet transform to the lowpass subband of the previous stage yields coarser temporal resolutions of the input sequence. This allows for lower bit rates, but also affects the visual quality of the lowpass subband.
So far, the number of total decomposition levels is determined for the entire input sequence in advance. However, if the motion in the video sequence is strong or if abrupt scene changes occur, a further decomposition leads to a low-quality lowpass subband. Therefore, we propose a content adaptive wavelet transform, which locally adapts the depth of the decomposition to the content of the input sequence. Thereby, the visual quality of the lowpass subband is increased by up to $10.28$\,dB compared to a uniform wavelet transform with the same number of total decomposition levels, while the required rate is reduced by $1.06\%$ additionally.
\end{abstract}
\section{Introduction}
\label{sec:intro}

\begin{figure}[t]
	\centering
	\resizebox{0.48\textwidth}{!}{%
		\begin{tikzpicture}[scale=1.1, >=latex'] 
		\input{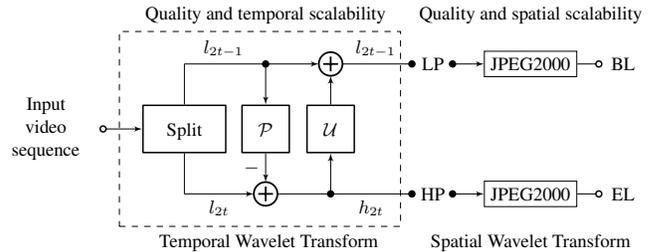}
		\tikzstyle{box} = [draw]	
		\coordinate (c0) at (-1.25,-1);
		\coordinate (c1) at (0,0);
		\coordinate (c2) at (0,-2);	
		\coordinate (c3) at (8.5,0);
		\coordinate (c4) at (8.5,-2);
		\node[left, align=center] (n0) at (-1.5,-1) {Input\\video\\sequence};
		\node[dspnodeopen,align=center] (n0) at (c0) {};
		\node[dspadder] (add1) at (2.25,0) {};
		\node[dspadder,dsp/label=below] (add3) at (1.25,-2) {};
		\node[dspsquare](b1) at (1.25,-1) {$\mathcal{P}$};
		\node[dspsquare](b2) at (2.25,-1) {$\mathcal{U}$};
		\node[dspfilter](split) at (0,-1) {Split};
		\node[dspnodefull](n5) at (1.25,0) {};
		\node[dspnodefull](n6) at (2.25,-2) {};
		\node (n9) at (3.825,0) {LP};
		\node (n10) at (3.825,-2) {HP};
		\node[dspnodefull] (n11) at (3.5,0) {};
		\node[dspnodefull] (n12) at (3.5,-2) {};
		\node[dspnodefull] (n13) at (4.125,0) {};
		\node[dspnodefull] (n14) at (4.125,-2) {};
		\node[box] (JPEG1) at (5.325,0) {JPEG2000};
		\node[box] (JPEG2) at (5.325,-2) {JPEG2000};
		\node (n15) at (6.75,0) {BL};
		\node (n16) at (6.75,-2) {EL};
		\node[dspnodeopen] (n17) at (6.375,0) {};
		\node[dspnodeopen] (n18) at (6.375,-2) {};
		\node (n19) at (1.3,-2.75) {Temporal Wavelet Transform};
		\node (n19) at (5.325,-2.75) {Spatial Wavelet Transform};
		\node (n20) at (1.25,0.75) {Quality and temporal scalability};
		\node (n21) at (5.325,0.75) {Quality and spatial scalability};
		\draw[-] (c1) -- node[above]{$l_{2t-1}$}(n5);
		\draw[->] (n5) -- (b1);
		\draw[->] (c2) -- node[below]{$l_{2t}$}(add3);
		\draw[->] (b1) -- node[left]{$-$}(add3);
		\draw[->] (n5) -- (add1);
		\draw[->] (b2) -- (add1);
		\draw[->] (n6) -- (b2);
		\draw[-] (add3) -- (n6);
		\draw[-] (add1) -- node[above]{$l_{2t-1}$}(n11);
		\draw[-] (n6) -- node[below]{$h_{2t}$}(n12);
		\draw[->] (n13) -- (JPEG1);
		\draw[-] (JPEG1) -- (n17);
		\draw[-] (JPEG2) -- (n18);
		\draw[->] (n14) -- (JPEG2);
		\draw[-] (split) -- (c1);
		\draw[-] (split) -- (c2);
		\draw[->] (n0) -- (split);
		\draw[dashed] (-1,-2.5) rectangle (3.325,0.5); 
		\end{tikzpicture}}
	\caption{Considered scenario to achieve a fully scalable representation of a video sequence. The dashed box shows the lifting structure of the wavelet transform for one decomposition level.}
	\label{fig:lifting}
\end{figure}

Many professional tasks like surveillance systems and telemedicine applications require lossless compression due to their sensitive content. However, lossless compression naturally leads to high bit rates. Considering any wireless network with limited channel capacity, a fast transmission of the entire data is challenging. Therefore, scalable lossless video coding is desirable, which allows for transmitting a base layer (BL) with coarser quality in the first instance and afterwards one or more enhancement layers (ELs), comprising the residual video data, to reconstruct the original sequence without any loss.
Basically, three different types of video scalability can be distinguished. Temporal scalability affects the frame rate, spatial scalability controls the spatial resolution, and quality scalability manipulates the fidelity of the video. 
Beside \mbox{DCT-based} coding schemes like Scalable High Efficiency Video Coding
(SHVC)~\cite{7172510} and Sample-Based Weighted Prediction for Enhancement Layer Coding (SELC)~\cite{lnt2017-42}, also 3-D subband coding (SBC)~\cite{Karlsson1988} can be applied. 3-D SBC is based on Wavelet Transforms (WT), which naturally provide scalability features without additional overhead~\cite{lnt2011-23}. As shown in Fig.\,\ref{fig:lifting}, by a transformation in temporal direction, the signal is decomposed into a lowpass (LP) and a highpass (HP) subband.
Both subbands offer only half the frame rate compared to the original sequence. While the LP subband is very similar to the original signal, the HP subband contains the structural information of the video sequence.
Afterwards, every frame of each subband is coded by the wavelet-based coder JPEG\,2000~\cite{ITU-T-800}, resulting in a fully scalable BL-EL-representation.

In this work, we focus on the optimization of the temporal scalability, which is controlled by the temporal WT highlighted by the dashed box in Fig.\,\ref{fig:lifting}.
The recursive application of the WT to the LP subband of the previous stage halves the frame rate for every decomposition level. 
This is advantageous for similar frames of the video sequence. In contrast, if huge changes occur among subsequent frames, the visual quality suffers significantly from multiple decomposition levels. This is why motion compensation (MC) should be included into the WT. 
However, MC always leads to a higher entire rate, mainly caused by the motion information, which has to be transmitted as additional overhead~\cite{8066388}. Further, there exists no practical approach for perfect MC. Hence, the error propagation will increase for a higher number of decomposition levels, leading to an inferior visual quality of the LP subband.
Therefore, the temporal scaling should be adapted to the video sequence. 
This can be reached by our proposed content adaptive wavelet lifting (CA-WL), which provides fine temporal resolution for high dynamic parts of a video sequence, while parts with few changes among subsequent frames are resolved coarser. 
After a brief overview of 3-D SBC, the proposed CA-WL is described in detail. Simulation results are given in the next section, followed by a short conclusion and outlook at the end of the paper.
\begin{figure*}
	\centering
	\resizebox{0.94\textwidth}{!}{%
		\begin{tikzpicture}[scale=0.5, >=latex'] 
		\input{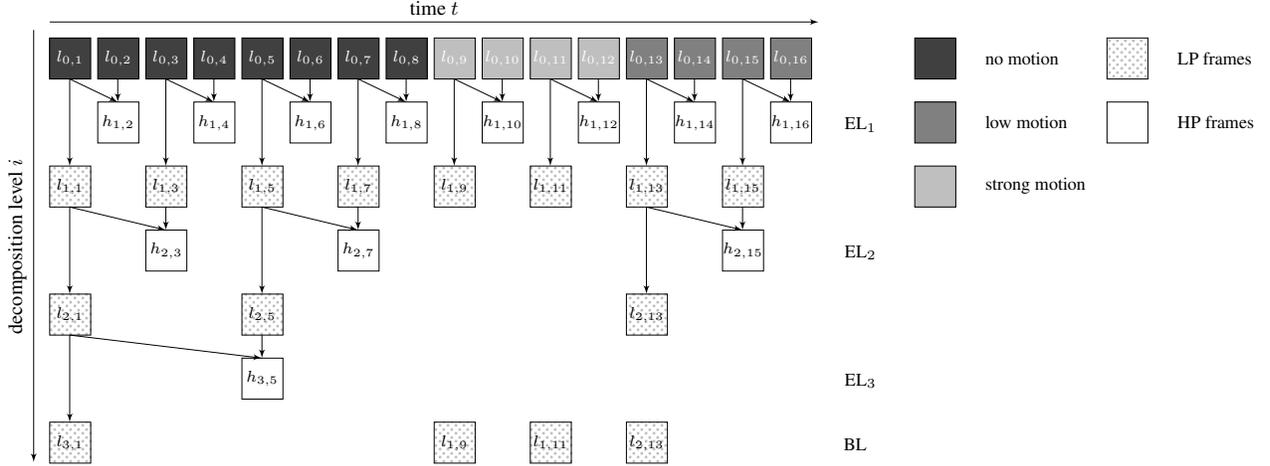}
		input{tikzlibrarydsp}
		\tikzstyle{box} = [draw]	
		\draw[->] (0,0.5) -- node[above]{time $t$}(24,0.5);	
		\draw[->] (-0.5,0.25) -- node[rotate=90,above]{decomposition level $i$}(-0.5,-13.25);
		\footnotesize
		\begin{scriptsize}	
		\foreach \x [count = \i] in {0,1.5,...,11}{
			\draw[fill=darkgray] (\x,-1.28) rectangle (\x+1.28,0) node[midway] {\textcolor{white}{$l_{0,\i}$}};
			\pgfmathparse{Mod(\i,2)==0?1:0}	
			\ifthenelse{\pgfmathresult>0}{	
				\draw[->] (\x+0.625,-1.28) -- (\x+0.625,-2);}{
				\draw[->] (\x+0.625,-1.28) -- (\x+0.625,-4);
				\draw[->] (\x+0.625,-1.28) -- (\x+2.125,-2);		}	}
		\foreach \x [count = \i] in {12,13.5,...,17}{
			\draw[fill=lightgray] (\x,-1.28) rectangle (\x+1.28,0) node[midway] {\textcolor{white}{$l_{0,\pgfmathparse{\i+8}\pgfmathprintnumber{\pgfmathresult}}$}};
			\pgfmathparse{Mod(\i,2)==0?1:0}	
			\ifthenelse{\pgfmathresult>0}{	
				\draw[->] (\x+0.625,-1.28) -- (\x+0.625,-2);}{
				\draw[->] (\x+0.625,-1.28) -- (\x+0.625,-4);
				\draw[->] (\x+0.625,-1.28) -- (\x+2.125,-2);		}	}
		\foreach \x [count = \i] in {18,19.5,...,23}{
			\draw[fill=gray] (\x,-1.28) rectangle (\x+1.28,0) node[midway] {\textcolor{white}{$l_{0,\pgfmathparse{\i+12}\pgfmathprintnumber{\pgfmathresult}}$}};
			\pgfmathparse{Mod(\i,2)==0?1:0}	
			\ifthenelse{\pgfmathresult>0}{	
				\draw[->] (\x+0.625,-1.28) -- (\x+0.625,-2);}{
				\draw[->] (\x+0.625,-1.28) -- (\x+0.625,-4);
				\draw[->] (\x+0.625,-1.28) -- (\x+2.125,-2);		}	}
		\foreach \x [count = \i] in {1.5,4.5,...,23}		
		\draw(\x,-3.28) rectangle (\x+1.28,-3.28+1.28) node [midway]{{$h_{1,\pgfmathparse{2+(\i-1)*2}\pgfmathprintnumber{\pgfmathresult}}$}};
		\foreach \x [count = \i] in {0,3,6,9}{
			\draw[pattern color=lightgray,pattern=crosshatch dots](\x,-5.28) rectangle (\x+1.28,-5.28+1.28) node [midway]{{$l_{1,\pgfmathparse{1+(\i-1)*2}\pgfmathprintnumber{\pgfmathresult}}$}};
			\pgfmathparse{Mod(\i,2)==0?1:0}	
			\ifthenelse{\pgfmathresult>0}{	
				\draw[->] (\x+0.625,-5.28) -- (\x+0.625,-6);}{
				\draw[->] (\x+0.625,-5.28) -- (\x+0.625,-8);
				\draw[->] (\x+0.625,-5.28) -- (\x+3.625,-6);		}	}
		\foreach \x [count = \i] in {12,15}{
			\draw[pattern color=lightgray,pattern=crosshatch dots](\x,-5.28) rectangle (\x+1.28,-5.28+1.28) node [midway]{{$l_{1,\pgfmathparse{9+(\i-1)*2}\pgfmathprintnumber{\pgfmathresult}}$}};	}	
		\foreach \x [count = \i] in {18,21}{
			\draw[pattern color=lightgray,pattern=crosshatch dots](\x,-5.28) rectangle (\x+1.28,-5.28+1.28) node [midway]{{$l_{1,\pgfmathparse{13+(\i-1)*2}\pgfmathprintnumber{\pgfmathresult}}$}};
			\pgfmathparse{Mod(\i,2)==0?1:0}	
			\ifthenelse{\pgfmathresult>0}{	
				\draw[->] (\x+0.625,-5.28) -- (\x+0.625,-6);}{
				\draw[->] (\x+0.625,-5.28) -- (\x+0.625,-8);
				\draw[->] (\x+0.625,-5.28) -- (\x+3.625,-6);		}	}
		\foreach \x [count = \i] in {3,9}
		\draw(\x,-7.28) rectangle (\x+1.28,-7.28+1.28) node [midway]{{$h_{2,\pgfmathparse{3+(\i-1)*2^2}\pgfmathprintnumber{\pgfmathresult}}$}};	
		\foreach \x [count = \i] in {21}
		\draw(\x,-7.28) rectangle (\x+1.28,-7.28+1.28) node [midway]{{$h_{2,15}$}};	
		\foreach \x [count = \i] in {0,6}{
			\draw[pattern color=lightgray,pattern=crosshatch dots](\x,-9.28) rectangle (\x+1.28,-9.28+1.28) node [midway]{{$l_{2,\pgfmathparse{1+(\i-1)*2^2}\pgfmathprintnumber{\pgfmathresult}}$}};
			\pgfmathparse{Mod(\i,2)==0?1:0}	
			\ifthenelse{\pgfmathresult>0}{	
				\draw[->] (\x+0.625,-9.28) -- (\x+0.625,-10);}{
				\draw[->] (\x+0.625,-9.28) -- (\x+0.625,-12);
				\draw[->] (\x+0.625,-9.28) -- (\x+6.625,-10);		}	}
		\foreach \x [count = \i] in {18}{
			\draw[pattern color=lightgray,pattern=crosshatch dots](\x,-9.28) rectangle (\x+1.28,-9.28+1.28) node [midway]{{$l_{2,13}$}};}
		\foreach \x [count = \i] in {6}
		\draw(\x,-11.28) rectangle (\x+1.28,-11.28+1.28) node [midway]{{$h_{3,\pgfmathparse{5+(\i-1)*2^3}\pgfmathprintnumber{\pgfmathresult}}$}};
		\foreach \x [count = \i] in {0}
		\draw[pattern color=lightgray,pattern=crosshatch dots](\x,-13.28) rectangle (\x+1.28,-13.28+1.28) node [midway]{{$l_{3,\i}$}};
		\draw[pattern color=lightgray,pattern=crosshatch dots] (12,-13.28) rectangle (12+1.28,-13.28+1.28) node [midway]{$l_{1,9}$};
		\draw[pattern color=lightgray,pattern=crosshatch dots] (15,-13.28) rectangle (15+1.28,-13.28+1.28) node [midway]{$l_{1,11}$};
		\draw[pattern color=lightgray,pattern=crosshatch dots] (18,-13.28) rectangle (18+1.28,-13.28+1.28) node [midway]{$l_{2,13}$};
		\end{scriptsize}
		\node[left] at (26,-2.7) {$\text{EL}_1$};
		\node[left] at (26,-6.7) {$\text{EL}_2$};
		\node[left] at (26,-10.7) {$\text{EL}_3$};
		\node[left] at (26,-12.7) {{$\text{BL}\hphantom{_1}$}};
		\draw[fill=darkgray] (27,-1.28) rectangle (27+1.28,0);
		\draw[fill=gray] (27,-3.28) rectangle (27+1.28,-3.28+1.28);
		\draw[fill=lightgray] (27,-5.28) rectangle (27+1.28,-5.28+1.28);
		\draw[pattern color=lightgray,pattern=crosshatch dots] (33,-1.28) rectangle (33+1.28,-1.28+1.28);
		\draw[fill=white] (33,-3.28) rectangle (33+1.28,-3.28+1.28);
		\node[right] (bl) at (29,-0.625) {no motion};
		\node[right] (gr) at (29,-2.625) {low motion};
		\node[right] (lg) at (29,-4.625) {strong motion};
		\node[right] (wh) at (35,-0.625) {LP frames};
		\node[right] (wh) at (35,-2.625) {HP frames};
		\end{tikzpicture}}
	\caption{Basic decomposition scheme of the CA-WL for $i_\text{max}{=}3$ decomposition levels resulting in one BL and three ELs. Depending on the underlying motion in the original sequence, the local depth of the decomposition differs.}
	\label{fig:model}
\end{figure*}

\section{3-D Subband Coding}
\label{sec:WT}

An efficient implementation of the discrete WT was proposed by Sweldens~\cite{Sweldens1995}. The \mbox{so-called} lifting structure consists of three steps: split, predict, and update. The block diagram of the lifting structure for a decomposition in temporal direction is depicted in the dashed box in Fig.\,\ref{fig:lifting}. In the first step, splitting is performed by decomposing the input signal into even- and odd-indexed frames $l_{2t}$ and $l_{2t-1}$. In the second step, the even frames are predicted from the odd frames by a prediction \mbox{operator $\mathcal{P}$}. Subtracting the predicted values $\mathcal{P}(l_{2t-1})$ from the even frames, results in the HP coefficients $h_{2t}$. In the third step, the HP coefficients are filtered by an update operator $\mathcal{U}$ and are added back to the odd frames, resulting in the LP coefficients. 
To get coarser temporal resolutions, the lifting scheme can be iterated on the LP subband by $i_{\text{max}}=\log_2(T)$ decomposition levels, where $T$ equals the total number of original frames. 

Since the lifting structure offers a flexible framework, it can be modified in multiple ways. By introducing rounding operators as introduced in~\cite{647983}, integer to integer transforms can be achieved, which yield perfect reconstruction. This makes the lifting structure of the WT highly attractive for many professional applications by offering a scalable representation and lossless reconstruction at the same time.
However, due to temporal displacements in the sequence, blurriness and ghosting artifacts will appear in the LP subband. 
These can be alleviated by incorporating MC methods directly into the lifting structure without losing the property of perfect reconstruction. This is called motion compensated temporal filtering (MCTF)~\cite{334985} and can be achieved by realizing the prediction \mbox{operator $\mathcal{P}$} by the warping operator $\mathcal{W}_{2t-1\rightarrow2t}$. Instead of the original odd frames, a compensated version is subtracted from the even frames. In case of the Haar wavelet filters, the prediction step is given by
\begin{align}
{h}_{2t} &= l_{2t}-\lfloor\mathcal{W}_{2t-1\rightarrow2t}(l_{2t-1})\rfloor. \label{eq:HP_mc}
\end{align}  
However, to achieve an equivalent wavelet transform, the compensation has to be inverted in the update step~\cite{bozinovic2005}. By reversing the index of $\mathcal{W}$, the LP coefficients of the Haar transform can be calculated by
\begin{equation}
{l}_{2t-1} = l_{2t-1}+\left\lfloor\frac{1}{2}\mathcal{W}_{2t\rightarrow2t-1}({h}_{2t})\right\rfloor. \label{eq:LP_mc}
\end{equation} 
$\mathcal{W}$ can be realized by different approaches of MC. In this work, we will employ a block-based approach.

\section{Content Adaptive Wavelet Lifting}
\label{sec:adaptiveWT}

Considering video sequences from surveillance systems or medical data sets, which comprise a temporal acquisition of images with contrast medium, there will be parts, where almost no motion occurs over time. In this case, an adaptive temporal scaling is advantageous, which performs iteratively a further decomposition, if subsequent frames are similar enough. If there are no changes over several frames, they shall be represented by only one LP frame. For significant changes among subsequent frames, for example when the contrast medium gets visible, these changes shall be represented in the LP subband with finer temporal resolution.

Fig.\,\ref{fig:model} shows the basic approach of our proposed content adaptive wavelet transform.
Index $i$ indicates the number of the current decomposition level. For $i{=}0$, no decomposition has been done so far, which corresponds to the original video sequence.   
In the first row, a schematic video sequence is given, which consists of three sections, each with a different amount of moving content. The corresponding amount of motion is described by the legend on the right side of Fig\,\ref{fig:model}. While in this example the first decomposition is performed for the entire sequence, the second decomposition is performed only on the frames with no or low motion. The third decomposition is exclusively done on frames with no motion. The resulting BL and ELs are marked at the right side. Since the maximum decomposition level $i_\text{max}$ is equal to $3$, three ELs are generated. By combining the ELs with the BL at the decoder side, the original sequence can be reconstructed step by step without any loss.

\subsection{Calculation of the Stopping Criterion}
\label{subsec:stoppingCrit}
Haar WTs can be represented with tree structures~\cite{lnt2011-23}. For 3-D SBC, the tree structure is given by decomposing two subsequent frames into LP and HP frames. 
To realize the CA-WL, the costs of the single nodes in every tree have to be considered. If the combined costs of the child nodes exceed the costs of the parent node, this means for an arbitrary signal $s$, if 
\begin{equation}
\mathcal{C}(s_{i,[2t-1,2t]}) \leq (\mathcal{C}(s_{i+1,2t-1})\cup\mathcal{C}(s_{i+1,2t}))
\end{equation}
holds, then the child nodes shall be pruned from the tree.
Here, $\mathcal{C}(\cdot)$ describes a cost functional, which represents the coding costs, such as entropy~\cite{119732} or rate-distortion~\cite{217221}. 
In this work, every decomposition level is performed for the entire input sequence in advance, before a retrospective evaluation of the resulting costs is done. Further, we decided to use a rate-distortion-based approach for calculating the coding costs. 
Therefore, we formulate the Lagrangian cost functional for a signal $s$ 
\begin{equation}
C(s) = D(s) + \lambda R(s).
\end{equation}
To determine the distortion $D(s)$ of the resulting LP frame, we calculate the MSE of the corresponding wavelet coefficients compared to the original signal according to~\cite{Lanz2018}. In this work, not only the similarity of the LP frame to the odd-indexed frame, but also the similarity to the even indexed frame is considered, which is a very important aspect for many professional applications.
The rate $R(s)$ is composed of the required rate for lossless coding of the LP and HP frames and, in case of MC, the file size of the motion vectors.
Then, the current decomposition can be evaluated locally by comparing the R-D costs of the children to the costs of the parent node for a given value $\lambda$. If the R-D costs of the child nodes exceed the costs of the parent node, thus if the following inequality
\begin{align*}
	D({l_{i,[2t-1,2t]}}) + \lambda R(l_{i,[2t-1,2t]}) \leq& \numberthis \\
	(D(l_{i+1,2t-1})+D(h_{i+1,2t})) + \lambda (R(l_{i+1,2t-1})&+R(h_{i+1,2t}))
\end{align*} 
holds, then a further decomposition is prevented. The Lagrange multiplier $\lambda$ can be any positive value. 
By choosing large values for $\lambda$, the rate is decreased, while for small values the distortion is decreased.

\begin{table}[t]
	\centering
	\caption{For the evaluation, sequences from surveillance systems, medical applications, and the HEVC test data set are employed. All sequences are used in 4:0:0 color sub-sampling format.}
	\label{tab:data}
	\resizebox{0.48\textwidth}{!}{%
		\begin{tabular}{l|l|ll}
			&& \textbf{Spatial resolution} & \textbf{Number of frames} \\ \hline
			\multirow{4}{*}{{\textbf{Surv}}}&\textit{AirportNight1}                 & $688\times352$              & $500$                     \\
			&\textit{AirportNight2}                 & $688\times432$              & $500$                     \\
			&\textit{AirportNight3}                 & $688\times372$              & $500$                     \\
			&\textit{AirportDay1}                   & $688\times432$              & $500$                     \\ \hline
			\multirow{2}{*}{{\textbf{Med}}}&\textit{MedFrontal}                          & $512\times512$              & $29$                      \\
			&\textit{MedSagittal}                          & $512\times512$              & $29$                      \\ \hline
			\multirow{2}{*}{{\textbf{HEVC}}}&\textit{ClassC}                          & $832\times480$              & $300$                     \\
			&\textit{ClassD}                          & $416\times240$              & $300$                    
		\end{tabular}
	}
\end{table}

\subsection{Handling of the Overhead}
For lossless reconstruction, the decomposition depth for every part of the input sequence has to be transmitted additionally. Therefore, a vector $\boldsymbol{v}$ is generated, whose length corresponds to $T$. This vector is initialized with zeros and gets an increment of $1$ at every temporal position of a LP frame after one decomposition level. The position to the corresponding HP frame is set to zero. Consequently, the non-zero entries correspond to the number of applied decomposition levels $i$ for every temporal position of a LP frame, while the distance $d$ to the corresponding HP frame is given by $d{=}2^{i-1}$. For the schematic video sequence in Fig.\,\ref{fig:model}, vector $\boldsymbol{v}$ is generated as follows:
\begin{align*}
&\text{Initialize } \boldsymbol{v}: & (0,0,0,0,0,0,0,0,0,0,0,0,0,0,0,0)	\\
&\boldsymbol{v} \text{ after level } i{=}1: & (1,0,1,0,1,0,1,0,1,0,1,0,1,0,1,0)\\
&\boldsymbol{v} \text{ after level } i{=}2: & (2,0,0,0,2,0,0,0,1,0,1,0,2,0,0,0)\\
&\boldsymbol{v} \text{ after level } i{=}3: & (3,0,0,0,0,0,0,0,1,0,1,0,2,0,0,0)
\end{align*}
The entire vector $\boldsymbol{v}$ is encoded using Context Adaptive Binary Arithmetic Coding (CABAC)~\cite{Witten:1987} and is transmitted to the decoder side.
Then, for lossless reconstruction of the previous stage, the decoder can easily determine the decomposition level and the temporal positions of the LP and HP frames regarding the original video sequence.


\section{Experimental Results}
\label{sec:results}

\begin{table}
	\centering
	\caption{Absolute $\text{PSNR}_{\text{LP}_t}$ [dB] and relative rate [\%] differences of our proposed CA-WL compared to the U-WL with (bottom) and without (top) block-based MC. Positive numbers denote a better visual quality and a higher rate of our proposed CA-WL and vice versa.}
	\resizebox{0.48\textwidth}{!}{%
		\begin{tabular}{l|l|l|c|c|c|c||c}
			&& $\boldsymbol{\lambda}$ & \textbf{Surv} &\textbf{Med} &\textbf{ClassC} &\textbf{ClassD} & \textbf{Total}\\
			&&  &  &  &  &  & \textbf{average}\\\hline
			\multirow{10}{*}{\tabrotate{No MC}}&\multirow{4}{*}{{$\Delta$ $\text{PSNR}_{\text{LP}_t}$}}  & 1 & 4.12 & 5.28 & 12.83 & 18.07 & $\bf{8.88}$  \\ 
			&& 3 & 1.64 & 1.91 & 6.32 & 11.4 & $\bf{5.3}$  \\ 
			&& 5 & 0.97 & 1.16 & 3.73 & 8.89 & $\bf{3.67}$  \\ 
			&& 7 & 0.65 & 1.16 & 4.09 & 8.27 & $\bf{3.5}$  \\ \cline{2-8} 
			&\multirow{4}{*}{{$\Delta$ File size}} & 1 & 5.99 & 0.09 & 11.64 & 9.07 & $\bf{6.56}$  \\ 
			&& 3 & 0.8 & -0.96 & 2.53 & 5.77 & $\bf{2.18}$  \\ 
			&& 5 & 0.23 & -1.29 & 0.84 & 4.04 & $\bf{1.08}$  \\ 
			&& 7 & 0.16 & -1.29 & 0.25 & 3.07 & $\bf{0.67}$  \\ \hhline{=|=|=|=|=|=|=#=}
			\multirow{13}{*}{\tabrotate{Block-based MC}}&\multirow{4}{*}{{$\Delta$ $\text{PSNR}_{\text{LP}_t}$}} & 1 & 9.3 & 15.56 & 6.99 & 14.15 & $\bf{10.98}$  \\ 
			&& 3 & 8.17 & 13.89 & 9.6 & 11.26 & $\bf{10.28}$  \\ 
			&& 5 & 7.42 & 13.89 & 9.21 & 9.55 & $\bf{9.47}$  \\ 
			&& 7 & 7.27 & 13.89 & 8.95 & 8.42 & $\bf{9.02}$  \\ \cline{2-8} 	
			&\multirow{4}{*}{{$\Delta$ File size}} & 1 & 0.16 & -5.58 & 1.98 & 6.9 & $\bf{1.34}$  \\ 
			&& 3 & -0.52 & -5.64 & -1.7 & 1.35 & $\bf{-1.06}$  \\ 
			&& 5 & -0.69 & -5.64 & -1.96 & 0.65 & $\bf{-1.38}$  \\ 
			&& 7 & -0.8 & -5.64 & -2.18 & 0.29 & $\bf{-1.57}$  \\ 	
		\end{tabular}
	}
	\label{tab:results_all}
\end{table}

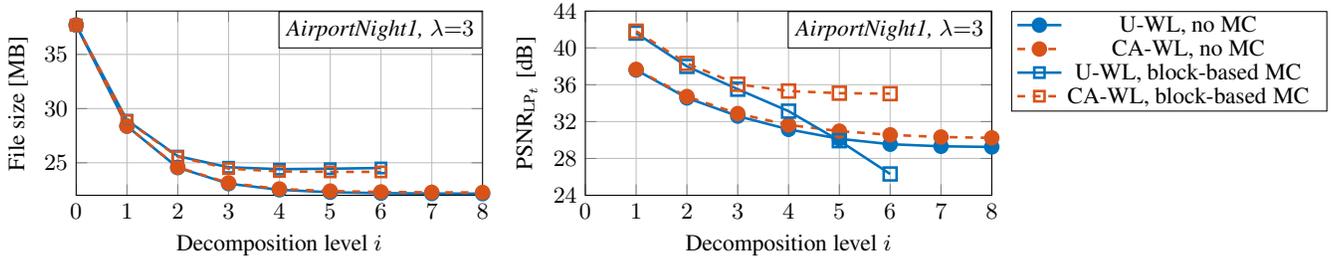
\begin{figure*}[tb]
	\centering 
	\resizebox{0.98\textwidth}{!}{%
	\begin{minipage}{\textwidth}
		\definecolor{mycolor1}{rgb}{0.00000,0.44700,0.74100}%
\definecolor{mycolor2}{rgb}{0.85000,0.32500,0.09800}%
\begin{tikzpicture}

\begin{groupplot}[group style={group name=myplot,group size= 2 by 1,horizontal sep=1.4cm, vertical sep=1.6cm},height=0.23\textwidth,width=0.4\textwidth]
	\nextgroupplot[
	xmin=0,
	xmax=8,
	xtick={0, 1, 2, 3, 4, 5, 6, 7, 8},
	xlabel={Decomposition level $i$},
	xmajorgrids,
	ymin=22,
	ymax=39,
	ylabel={File size [MB]},
	ymajorgrids,
	yminorgrids,
	axis background/.style={fill=white},
	every axis title/.style={draw,fill=white,below left,at={(1,1)}},
	title={\textit{AirportNight1, }$\lambda{=}3$}
	]
	\addplot [color=mycolor1,solid,line width=1.0pt,mark size=2.5pt,mark=*,mark options={solid}]
	table[row sep=crcr]{%
		0	37.7067403793335\\
		1	28.3827543258667\\
		2	24.5557012557983\\
		3	23.075701713562\\
		4	22.5063352584839\\
		5	22.2823076248169\\
		6	22.19797706604\\
		7	22.162956237793\\
		8	22.1506185531616\\
	};\label{size_uniform_none}
	\addplot [color=mycolor2,dashed,line width=1.0pt,mark size=2.5pt,mark=*,mark options={solid}]
	table[row sep=crcr]{%
		0	37.7067403793335\\
		1	28.3874597549438\\
		2	24.5866746902466\\
		3	23.1406936645508\\
		4	22.582857131958\\
		5	22.3897113800049\\
		6	22.3105058670044\\
		7	22.2739458084106\\
		8	22.261607170105\\
	};\label{size_adaptive_none}
    \addplot [color=mycolor1,solid,line width=1.0pt,mark size=2.0pt,mark=square,mark options={solid}]
    table[row sep=crcr]{%
    	0	37.7067403793335\\
    	1	28.9264507293701\\
    	2	25.6154565811157\\
    	3	24.5852661132812\\
    	4	24.4128932952881\\
    	5	24.4521389007568\\
    	6	24.5201025009155\\
    };\label{size_uniform_block}
    \addplot [color=mycolor2,dashed,line width=1.0pt,mark size=2.0pt,mark=square,mark options={solid}]
    table[row sep=crcr]{%
    	0	37.7067403793335\\
    	1	28.8917722702026\\
    	2	25.5542869567871\\
    	3	24.4621267318726\\
    	4	24.2029428482056\\
    	5	24.1628198623657\\
    	6	24.1582298278809\\
    };\label{size_adaptive_block}   
    \nextgroupplot[title style={font=\bfseries},
    xmin=0,
    xmax=8,
    xtick={0, 1, 2, 3, 4, 5, 6, 7, 8},
    xlabel={Decomposition level $i$},
    xmajorgrids,
    ymin=24,
    ymax=44,
    ytick={24,  28,  32,  36,  40,  44},
    ylabel={$\text{PSNR}_{\text{LP}_t}$ [dB]},
    ymajorgrids,
    yminorgrids,
    axis background/.style={fill=white},
    every axis title/.style={draw,fill=white,below left,at={(1,1)}},
    title={\textit{AirportNight1, }$\lambda{=}3$}
    ]
    \addplot [color=mycolor1,solid,line width=1.0pt,mark size=2.5pt,mark=*,mark options={solid}]
    table[row sep=crcr]{%
    	1	37.6030843339115\\
    	2	34.5736366456155\\
    	3	32.5973827901899\\
    	4	31.1566739920551\\
    	5	30.1336046208568\\
    	6	29.5454768706041\\
    	7	29.3192443584658\\
    	8	29.2484095120996\\
    };\label{psnr_uniform_none}
    \addplot [color=mycolor2,dashed,line width=1.0pt,mark size=2.5pt,mark=*,mark options={solid}]
    table[row sep=crcr]{%
    	1	37.6665918286883\\
    	2	34.7215187163142\\
    	3	32.8714811866963\\
    	4	31.637210967376\\
    	5	30.9672460305921\\
    	6	30.5613084130767\\
    	7	30.3251095035174\\
    	8	30.2360014399652\\
    };\label{psnr_adaptive_none}
    \addplot [color=mycolor1,solid,line width=1.0pt,mark size=2.5pt,mark=square,mark options={solid}]
    table[row sep=crcr]{%
    	1	41.6036950623889\\
    	2	37.9903627521675\\
    	3	35.5161961725151\\
    	4	33.1347236401194\\
    	5	29.9282862094466\\
    	6	26.3142802370222\\
    };\label{psnr_uniform_block}    
    \addplot [color=mycolor2,dashed,line width=1.0pt,mark size=2.5pt,mark=square,mark options={solid}]
    table[row sep=crcr]{%
    	1	41.8146676928222\\
    	2	38.3172161509547\\
    	3	36.0592079730663\\
    	4	35.3173464494491\\
    	5	35.0837547337752\\
    	6	35.0371071813401\\
    };\label{psnr_adaptive_block}          
\end{groupplot}
 legend
\path (myplot c1r1.south east|-current bounding box.south)--
      coordinate(legendpos)
      (myplot c2r1.south west |-current bounding box.south);
\matrix[
    matrix of nodes,
    anchor=west,
    draw,
    inner sep=0.1em,
    draw
  ]at([xshift=6.5cm,yshift=2.75cm]legendpos)
  {
    \ref{size_uniform_none}& U-WL, no MC&[5pt]\\
    \ref{size_adaptive_none}& CA-WL, no MC&[5pt]\\
    \ref{size_uniform_block}& U-WL, block-based MC&[5pt]\\
    \ref{size_adaptive_block}& CA-WL, block-based MC&[5pt]\\};

\end{tikzpicture}
	\end{minipage}
	}
	\vspace{-0.5cm}
	\caption{Absolute rate and $\text{PSNR}_{\text{LP}_t}$ results from the \textit{AirportNight1} sequence with and without MC, using $\lambda{=}3$. The results are displayed over all reached decomposition levels $i$. The proposed method is characterized by the dashed lines.}
	\label{fig:RD_results} 
	\vspace{0.15cm}
\end{figure*} 

\begin{figure*}[t!]
	\centering
	\resizebox{\textwidth}{!}{%
		\begin{tikzpicture}
		\node (U-WL) at (4.6,1.7) {\textbf{U-WL}};
		\node (CA-WL) at (9.2,1.7) {\textbf{Proposed CA-WL}};
		\node (U-WL) at (13.8,1.7) {\textbf{U-WL}};
		\node (CA-WL) at (0.2,1.7) {\textbf{Original}};
		\node (CA-WL) at (18.4,1.7) {\textbf{Proposed CA-WL}};
		\node (U-WL) at (6.9,2.2) {\textbf{No MC}};
		\node (CA-WL) at (16.1,2.2) {\textbf{Block-based MC}};
		\draw[-,dashed] (2.3,2.4) -- (2.3,-6.75);
		\draw[-,dashed] (11.5,2.4) -- (11.5,-6.75);
		%
		\node[align=center, rotate=90,] (surf) at (-2.8,0) {\small{\textbf{\textit{AirportDay1}}}\\ \small{Frame $321$}};
		%
		\node[inner sep=0pt] (orig_surf) at (0,0)
			{\includegraphics[trim = 2.5cm 1cm 2.5cm 0.75cm,clip,width=.25\textwidth]{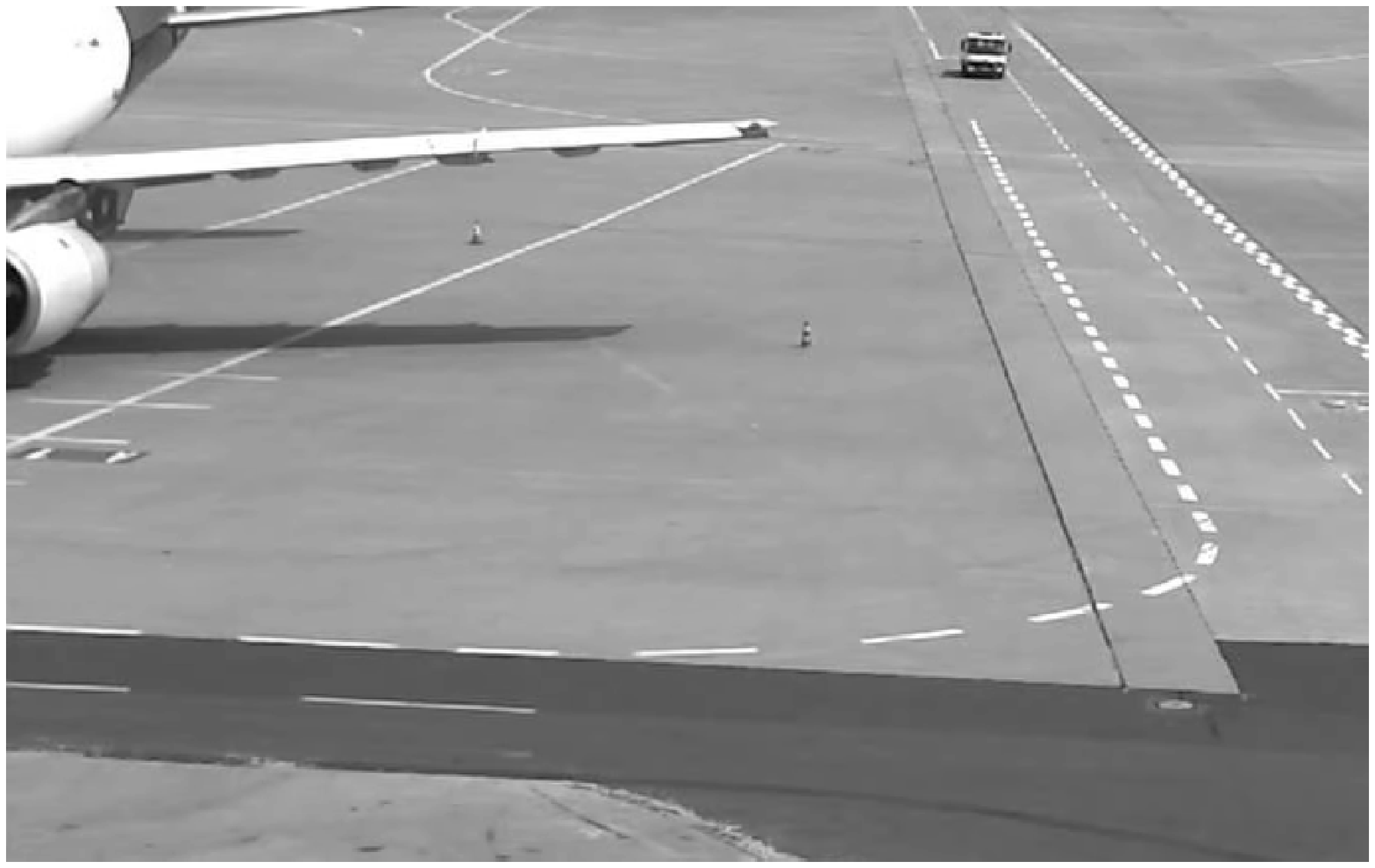}};		
		\node[inner sep=0pt] (UWT_none) at (4.6,0)
			{\includegraphics[trim = 2.5cm 1cm 2.5cm 0.75cm,clip,width=.25\textwidth]{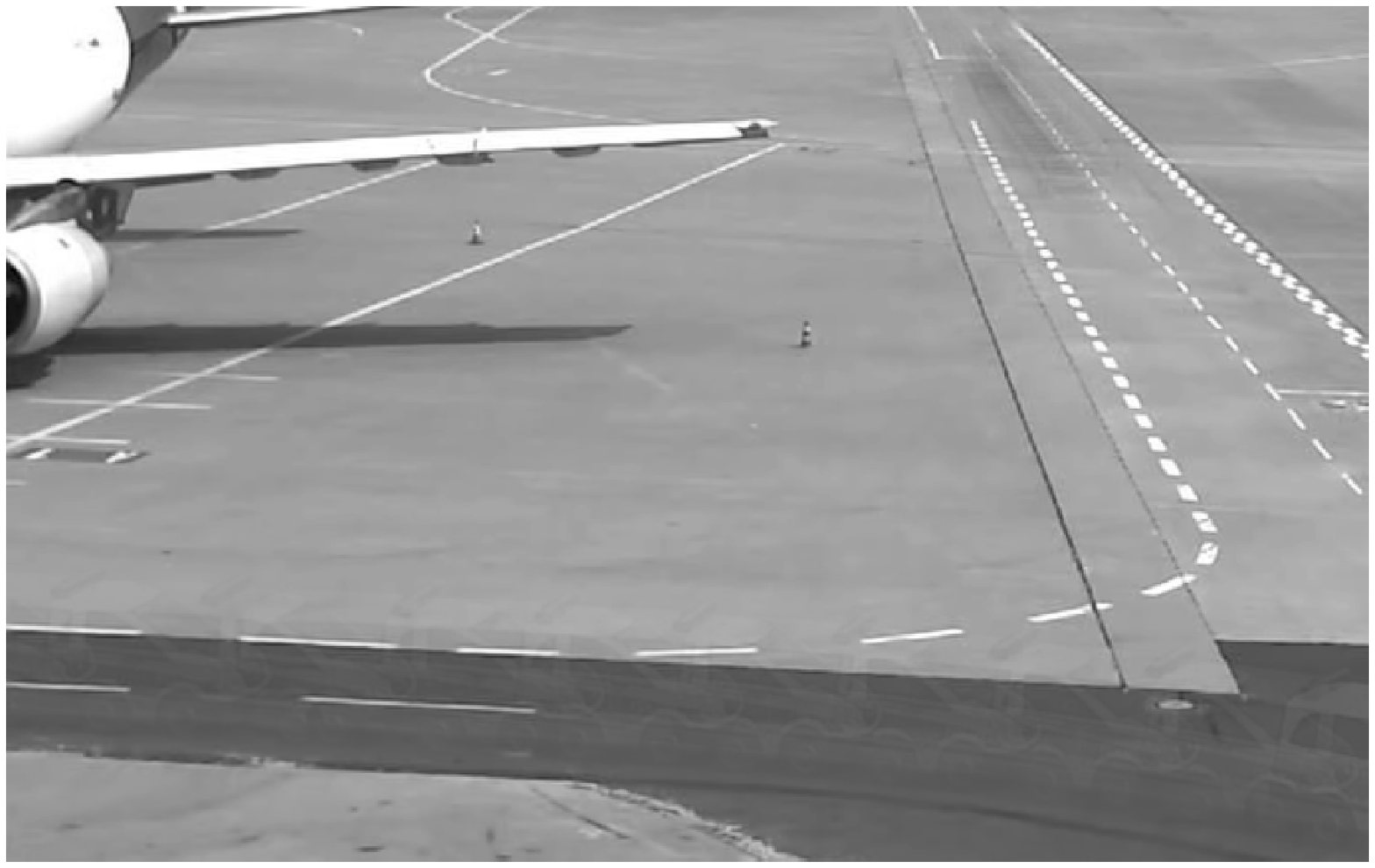}};
			\node[fill=white] (n1) at (6.15,-1.15) {\small{$35.01$\,dB}};
			\draw[draw=red,very  thick]  (5.65,1.2) ellipse (0.2cm and 0.2cm);	
		\node[inner sep=0pt] (CAWT_none) at (9.2,0)
			{\includegraphics[trim = 2.5cm 1cm 2.5cm 0.75cm,clip,width=.25\textwidth]{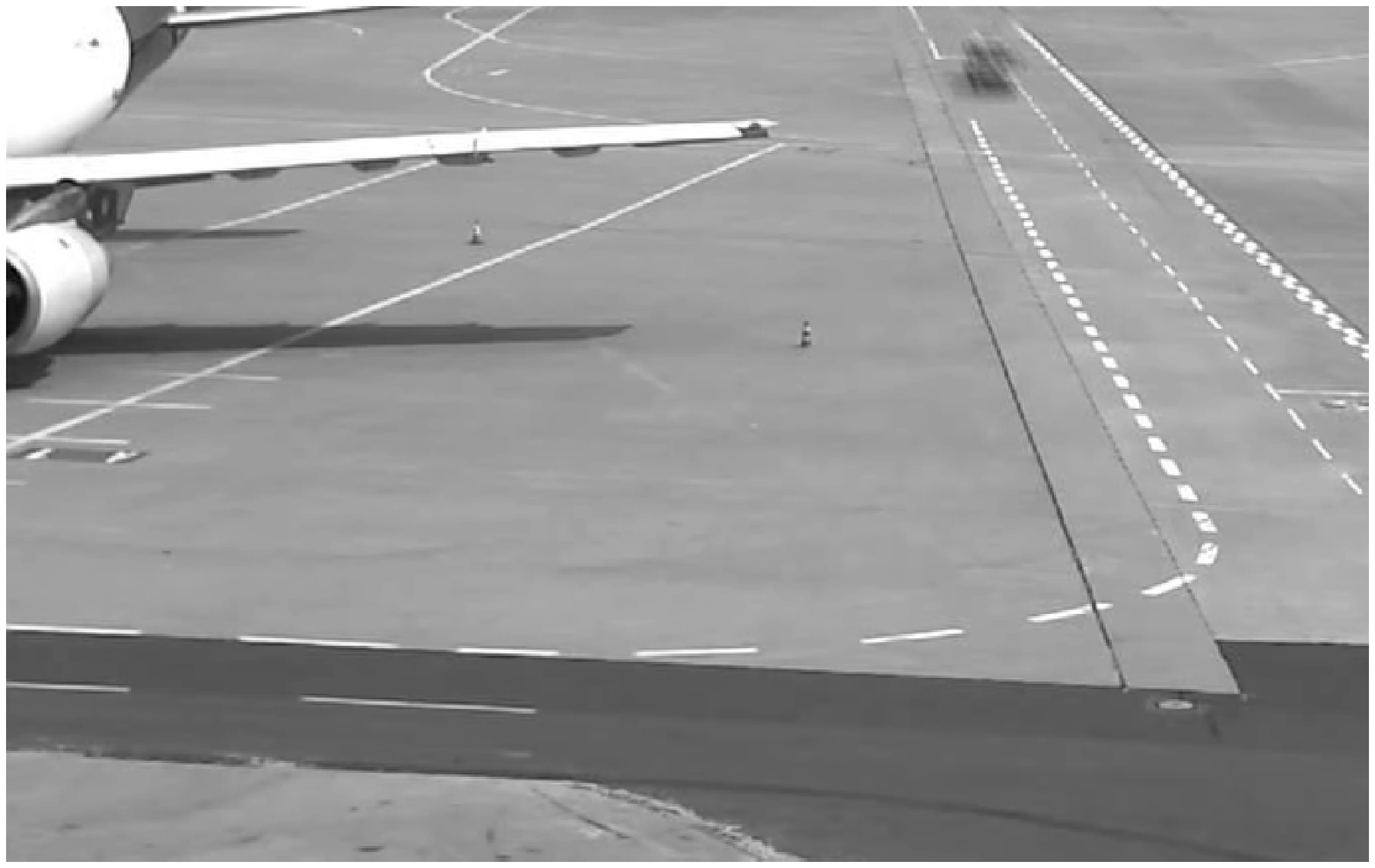}};
			\node[fill=white] (n1) at (10.75,-1.15) {\small{$37.28$\,dB}};			
		\node[inner sep=0pt] (UWT_block) at (13.8,0)
			{\includegraphics[trim = 2.5cm 1cm 2.5cm 0.75cm,clip,width=.25\textwidth]{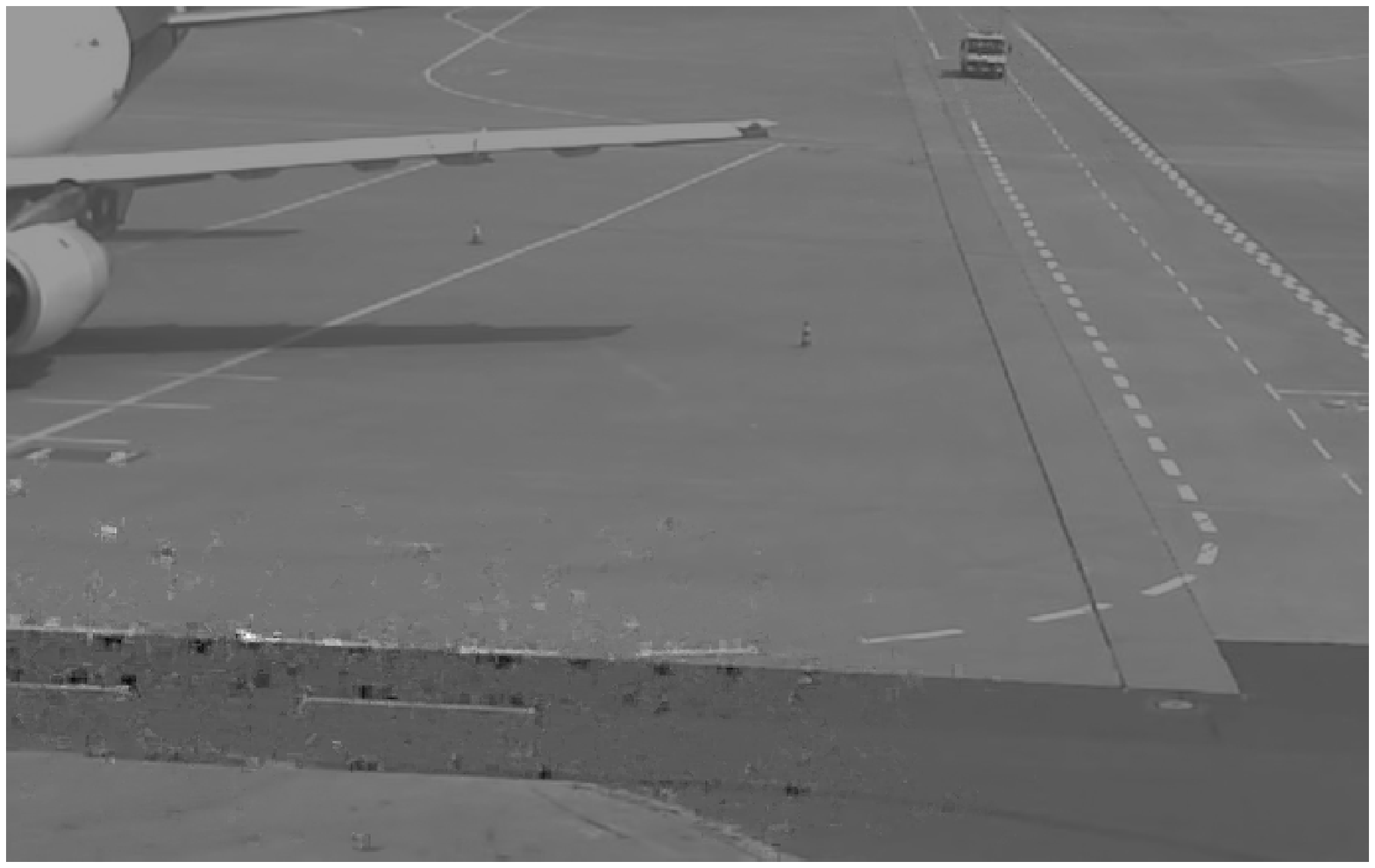}};
			\node[fill=white] (n1) at (15.35,-1.15) {\small{$30.75$\,dB}};
			\draw[draw=red,very  thick]  (11.7,-1.2) rectangle (11.7+2.5,-1.2+1);
		\node[inner sep=0pt] (CAWT_block) at (18.4,0)
			{\includegraphics[trim = 2.5cm 1cm 2.5cm 0.75cm,clip,width=.25\textwidth]{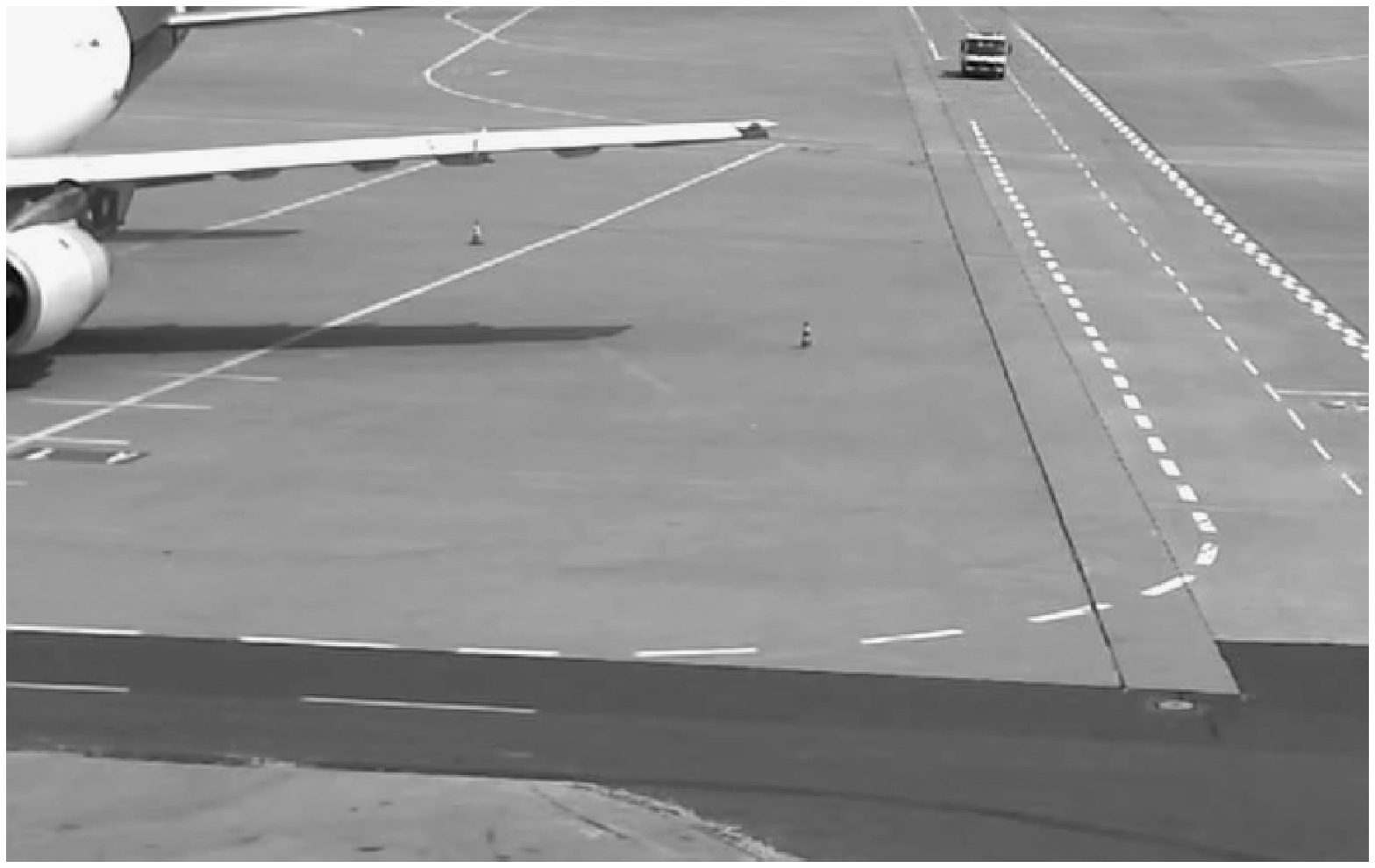}};
			\node[fill=white] (n1) at (19.95,-1.15) {\small{$47.02$\,dB}};
		\node[align=center, rotate=90,] (surf) at (-2.8,-2.85) {\small{\textbf{\textit{MedSagittal}}}\\ \small{Frame $17$},zoom};
		%
		\node[inner sep=0pt] (orig_surf_med_zoom) at (0,-2.85)
			{\includegraphics[trim = 4cm 4cm 8cm 8cm,clip,width=.25\textwidth]{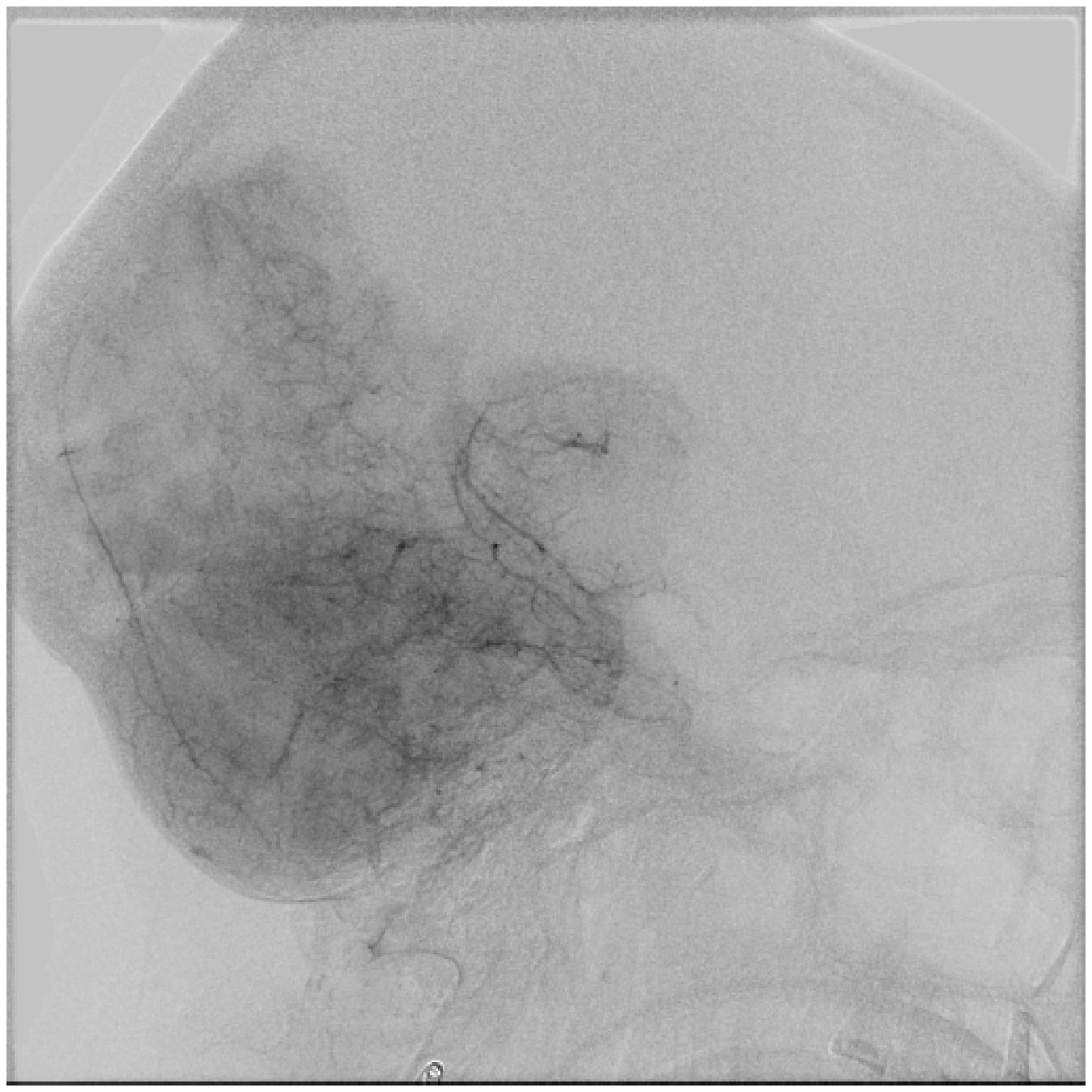}};
		\node[inner sep=0pt] (UWT_none) at (4.6,-2.85)
			{\includegraphics[trim = 4cm 4cm 8cm 8cm,clip,width=.25\textwidth]{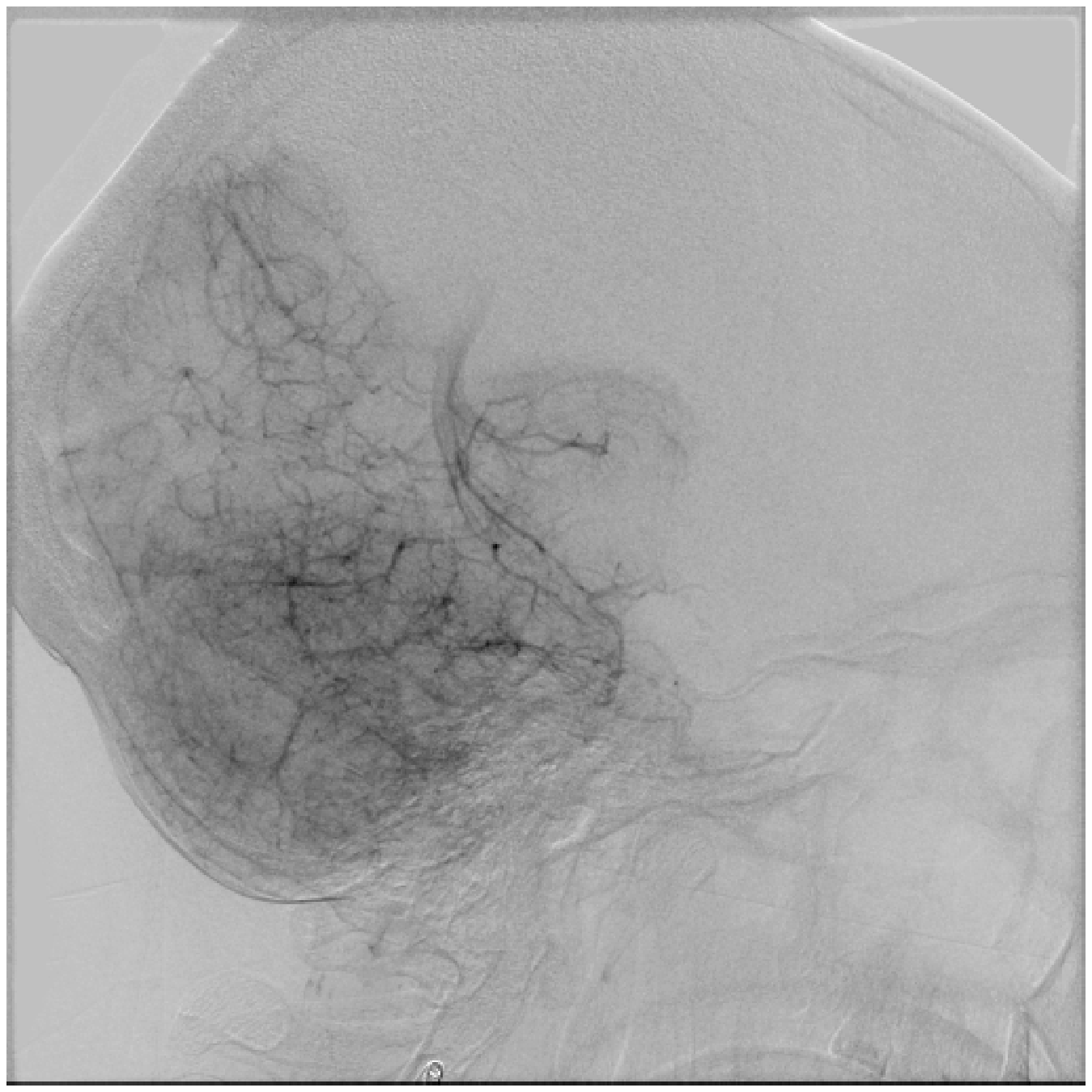}};
			\node[fill=white] (n1) at (6.15,-4) {\small{$39.67$\,dB}};
			\draw[draw=red,very  thick]  (5.5,-3.0) ellipse (0.5cm and 0.5cm);
		\node[inner sep=0pt] (CAWT_none) at (9.2,-2.85)
			{\includegraphics[trim = 4cm 4cm 8cm 8cm,clip,width=.25\textwidth]{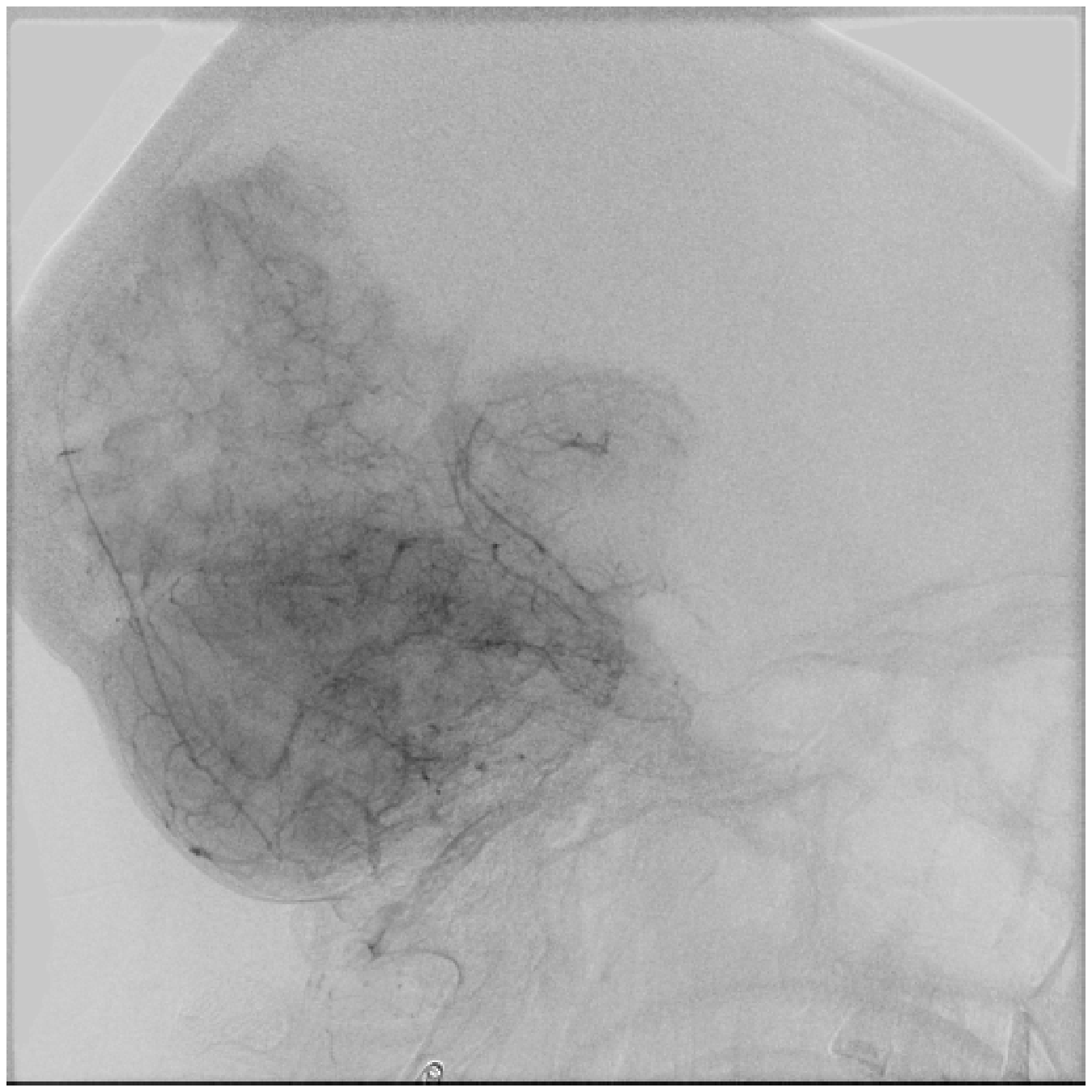}};
			\node[fill=white] (n1) at (10.75,-4) {\small{$44.34$\,dB}};		
		\node[inner sep=0pt] (UWT_block) at (13.8,-2.85)
			{\includegraphics[trim = 4cm 4cm 8cm 8cm,clip,width=.25\textwidth]{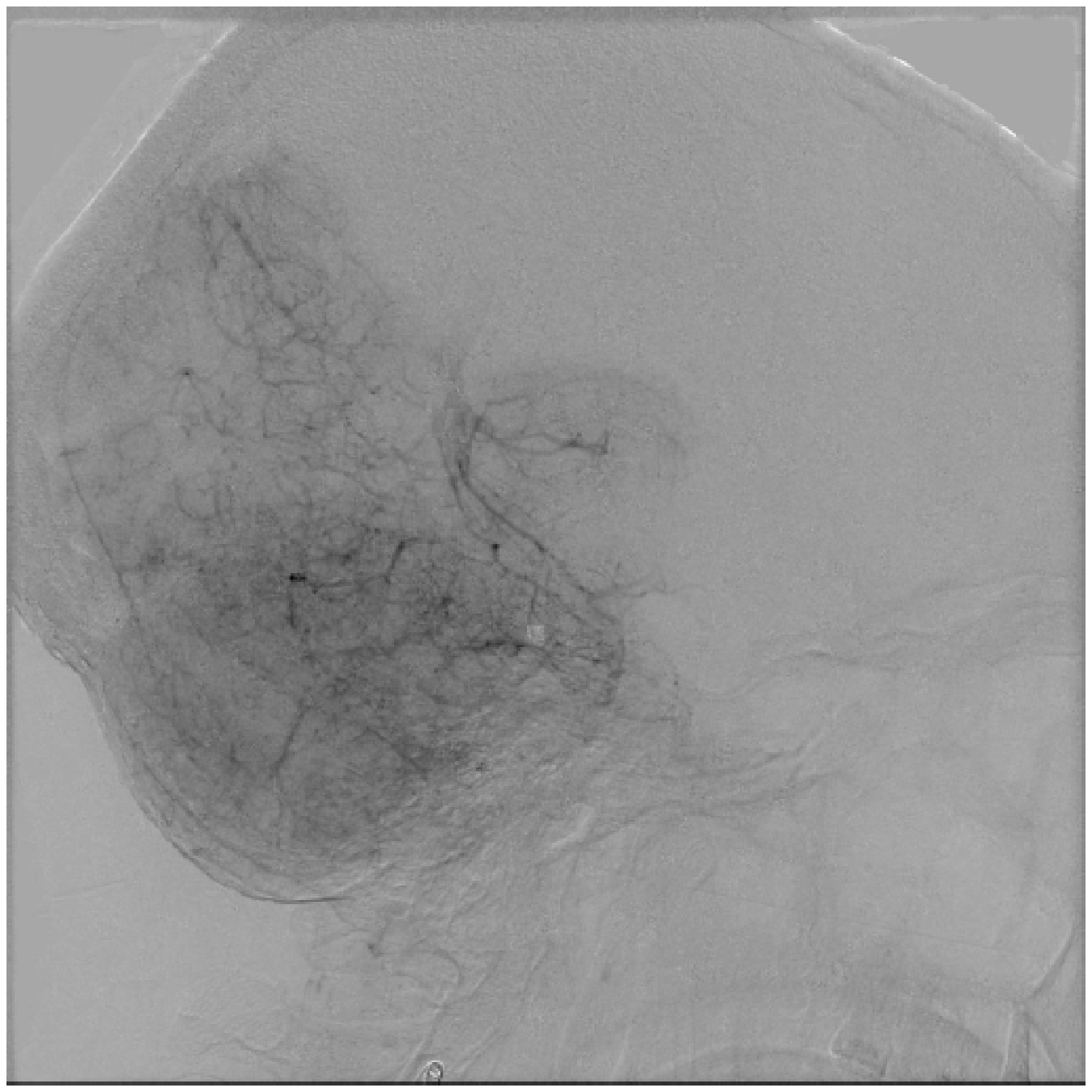}};
			\node[fill=white] (n1) at (15.35,-4) {\small{$39.46$\,dB}};
			\draw[draw=red,very  thick]  (14.75,-2.25) rectangle (14.75+0.6,-2.25+0.6);
		\node[inner sep=0pt] (CAWT_block) at (18.4,-2.85)
			{\includegraphics[trim = 4cm 4cm 8cm 8cm,clip,width=.25\textwidth]{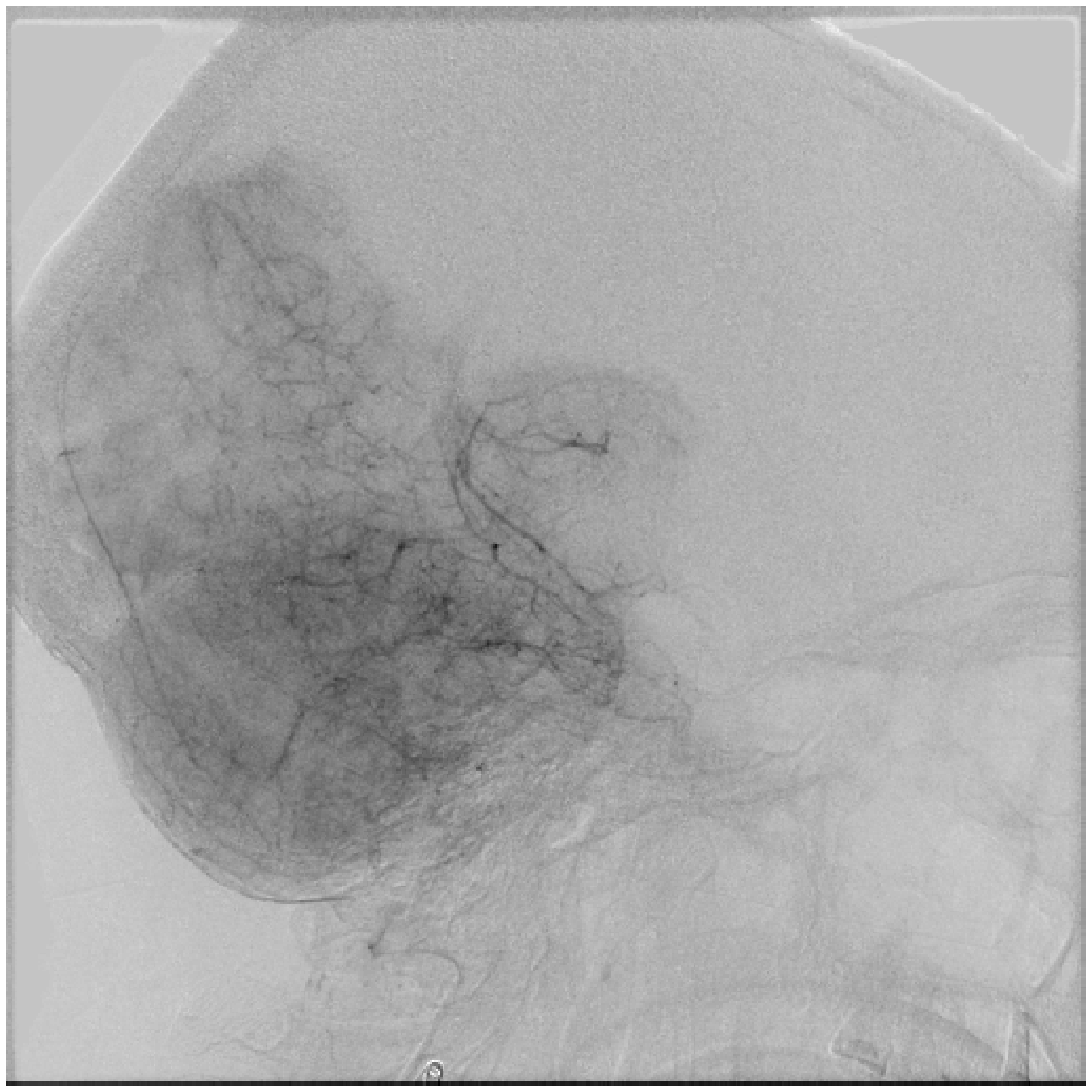}};
			\node[fill=white] (n1) at (19.95,-4) {\small{$43.92$\,dB}};
		\node[align=center, rotate=90,] (surf) at (-2.8,-5.7) {\small{\textbf{\textit{BQMall}}}\\ \small{Frame $25$, zoom}};
		%
		\node[inner sep=0pt] (orig_BQMall_zoom) at (0,-5.7)
		{\includegraphics[trim = 2.5cm 1cm 14cm 7.9cm,clip,width=.25\textwidth]{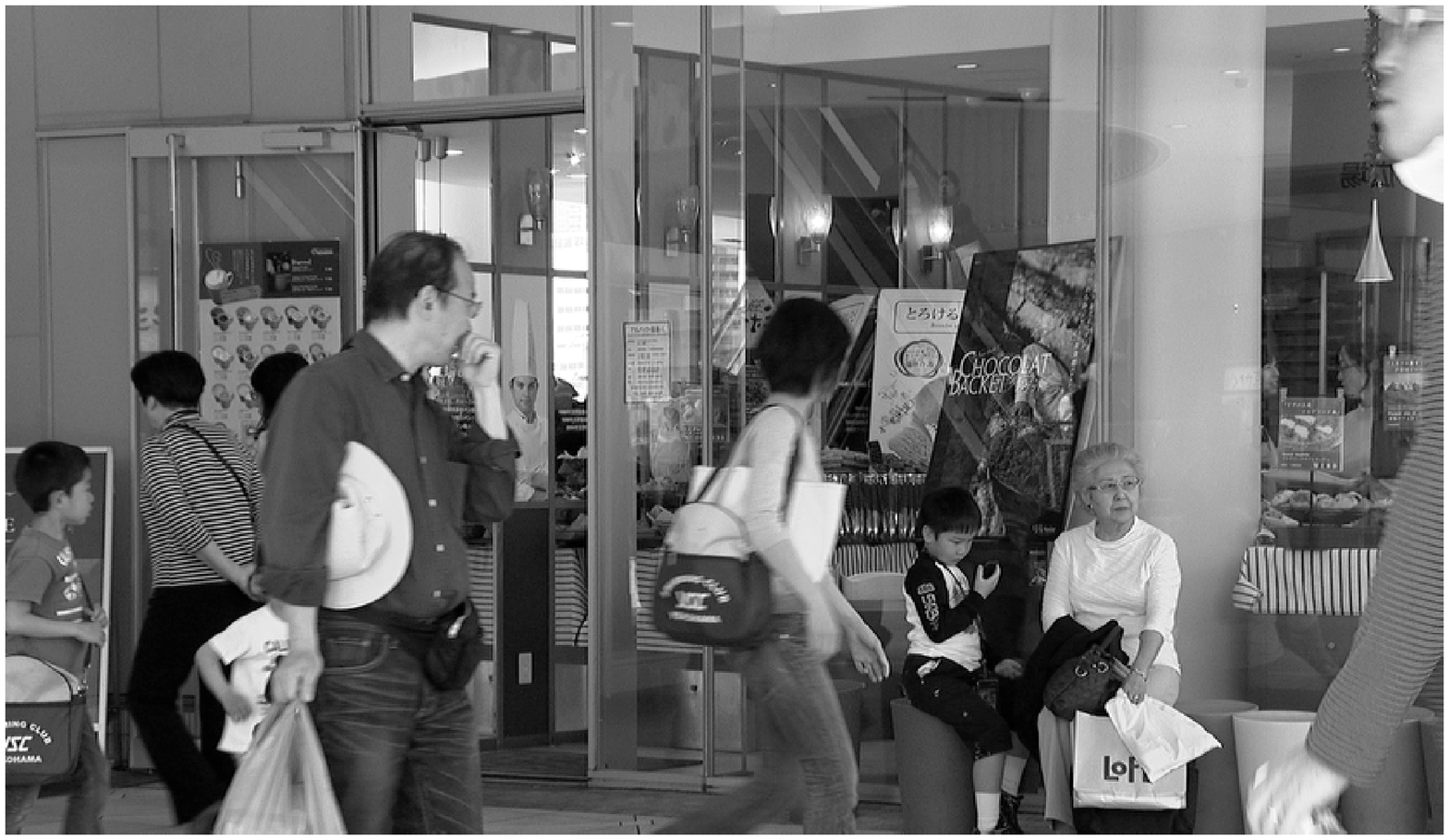}};
		\node[inner sep=0pt] (UWT_none) at (4.6,-5.7)
			{\includegraphics[trim = 2.5cm 1cm 14cm 7.9cm,clip,width=.25\textwidth]{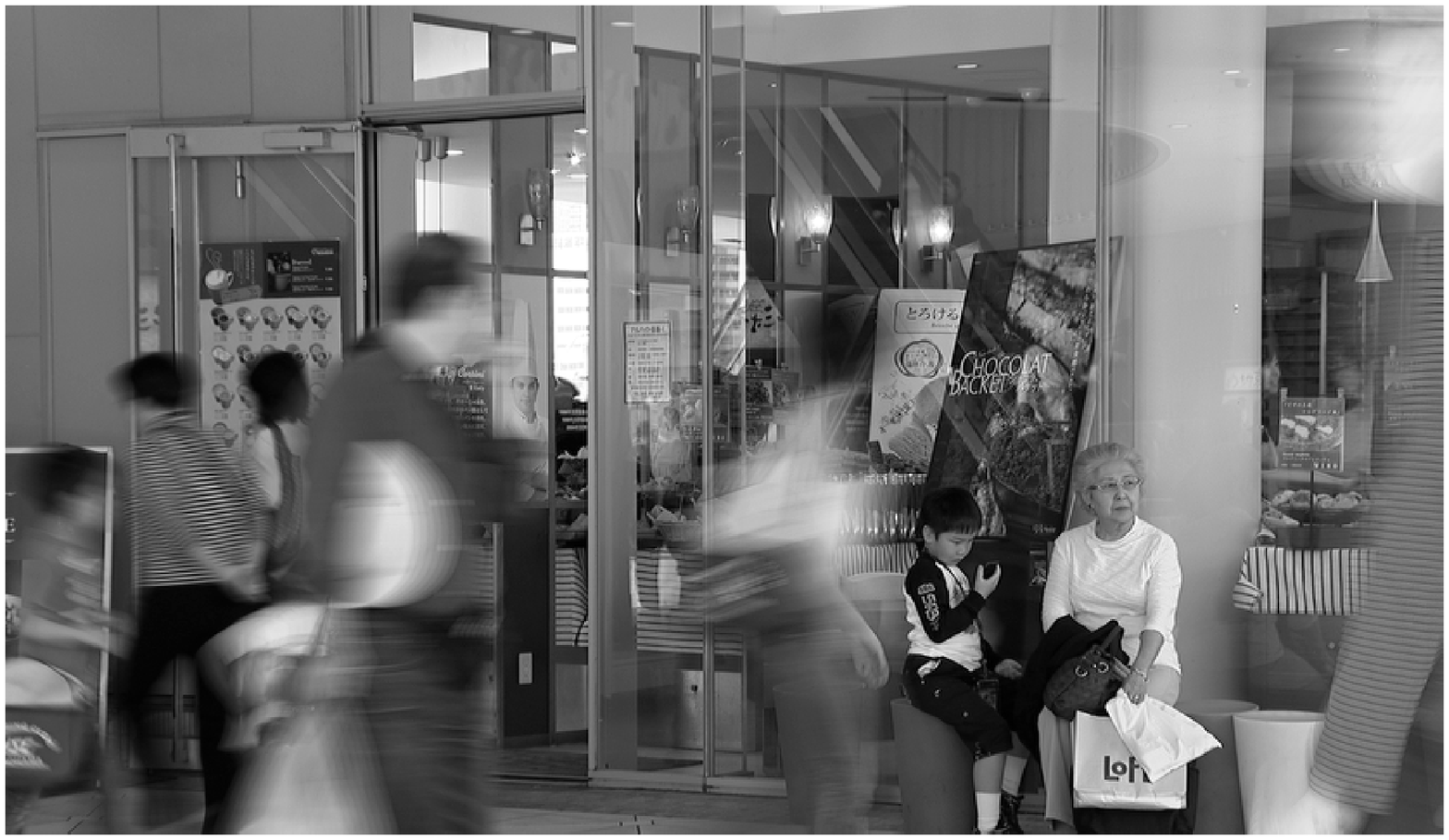}};
			\node[fill=white] (n1) at (6.15,-6.55) {\small{$20.87$\,dB}};
			\draw[rotate around ={-25:(4.5,-5.6)}, draw=red,very  thick]  (4.5,-5.6) ellipse (0.8cm and 1.15cm);
		\node[inner sep=0pt] (CAWT_none) at (9.2,-5.7)
			{\includegraphics[trim = 2.5cm 1cm 14cm 7.9cm,clip,width=.25\textwidth]{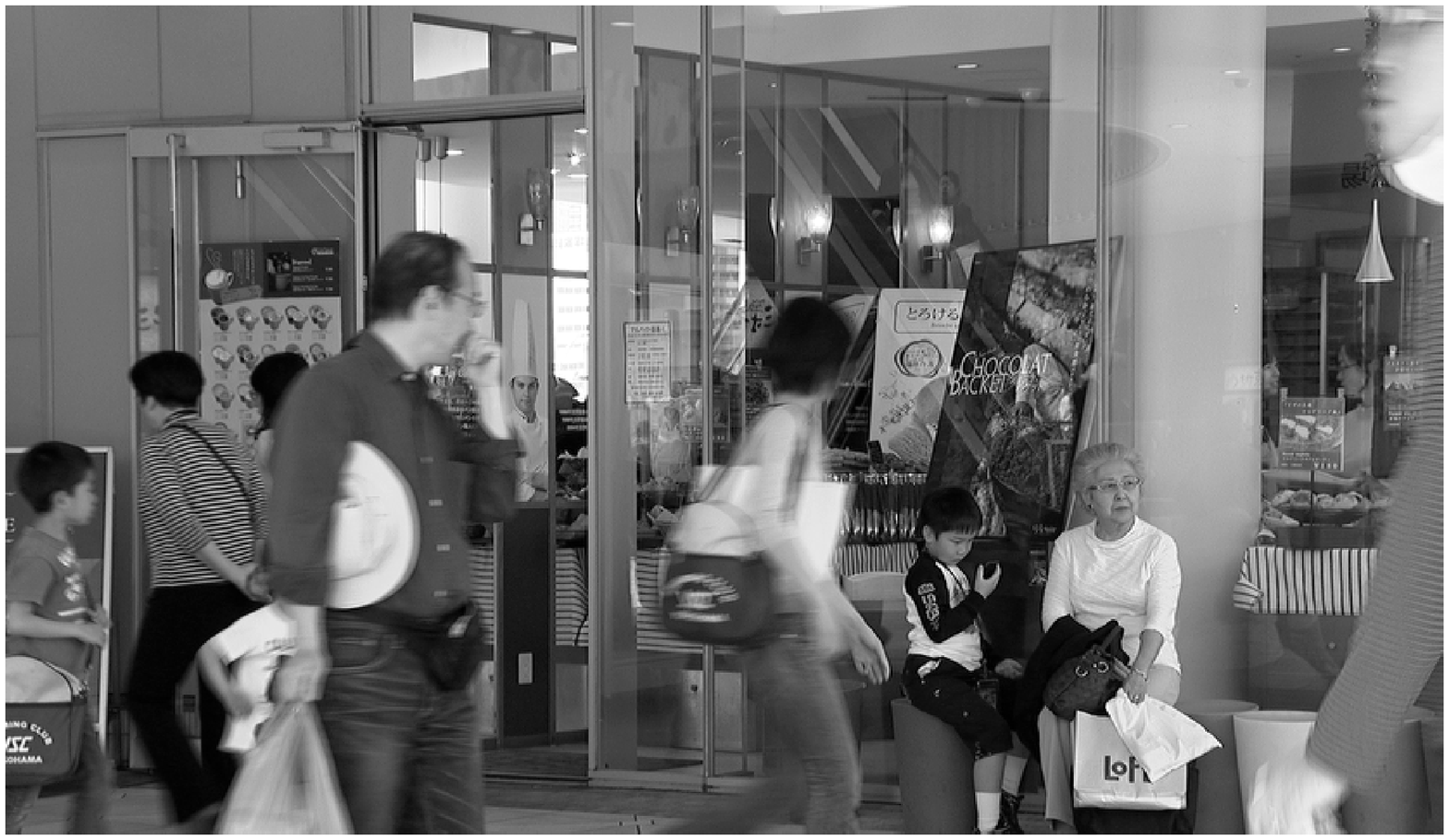}};
			\node[fill=white] (n1) at (10.75,-6.55) {\small{$29.69$\,dB}};		
		\node[inner sep=0pt] (UWT_block) at (13.8,-5.7)
			{\includegraphics[trim = 2.5cm 1cm 14cm 7.9cm,clip,width=.25\textwidth]{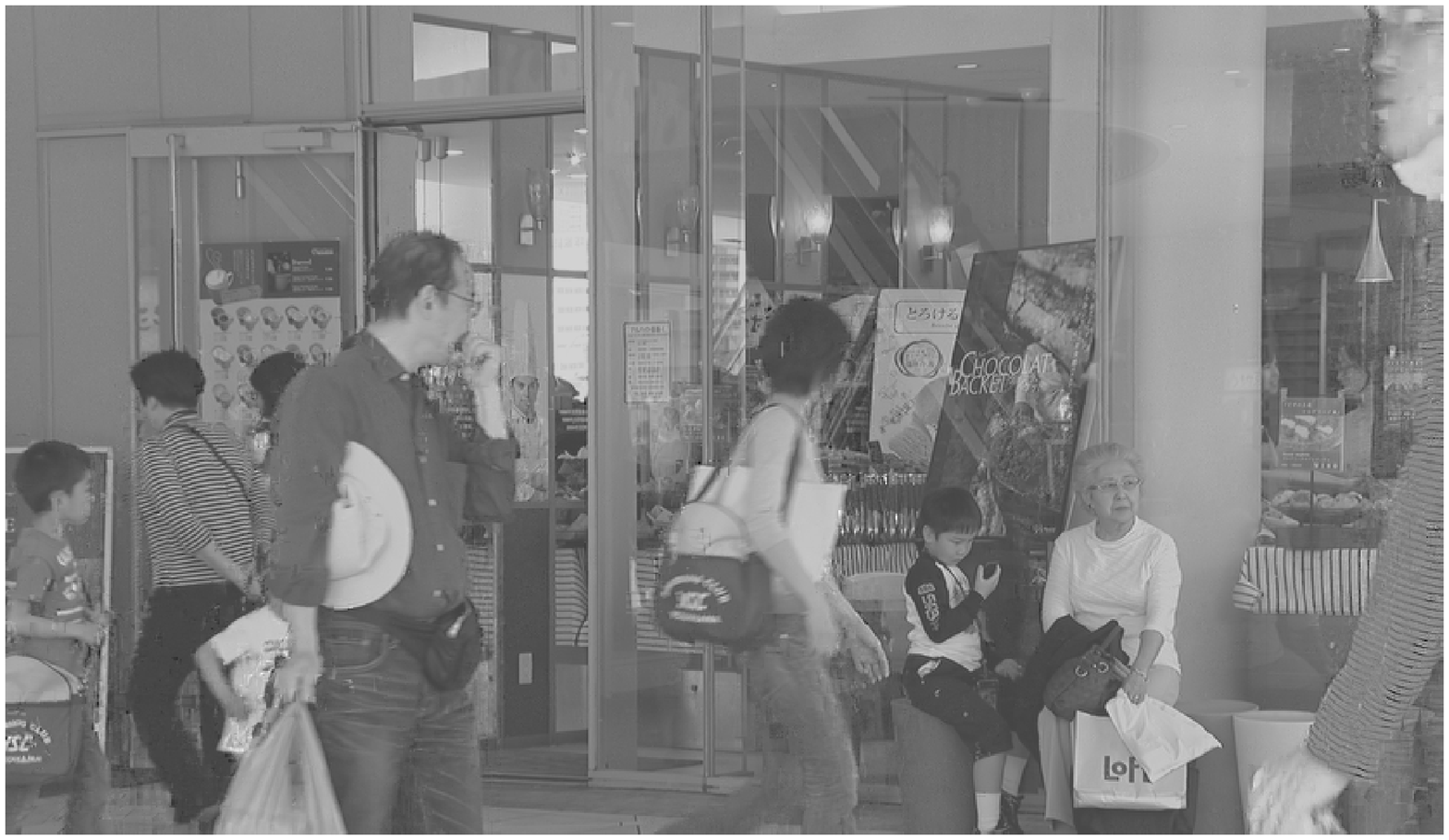}};
			\node[fill=white] (n1) at (15.35,-6.55) {\small{$30.23$\,dB}};
			\draw[draw=red,very  thick]  (11.8,-6.6) rectangle (11.8+1,-6.6+1.5);
		\node[inner sep=0pt] (CAWT_block) at (18.4,-5.7)
			{\includegraphics[trim = 2.5cm 1cm 14cm 7.9cm,clip,width=.25\textwidth]{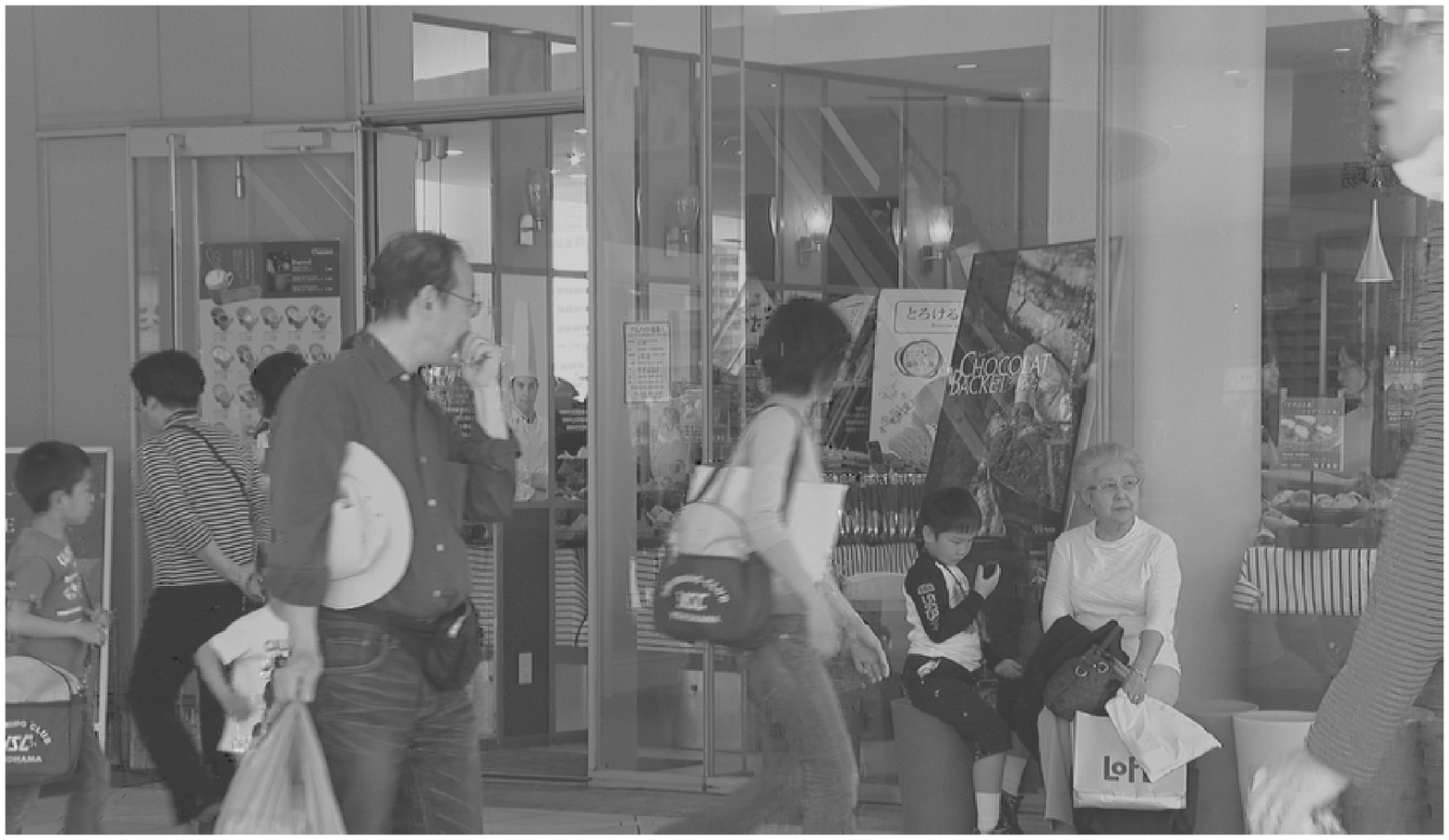}};
			\node[fill=white] (n1) at (19.95,-6.55) {\small{$33.54$\,dB}};
		\end{tikzpicture}}
	\caption{Comparison of the visual quality of one frame from each test data set compared to the corresponding LP frames of a U-WL and our CA-WL with and without block-based MC, for $\lambda=3$. The rectangles depict blocking artifacts, the circles indicate missing objects, and the ellipses show blurring artifacts.}
	\label{fig:visual_results}
\end{figure*}

Our simulation setup comprises surveillance videos, medical sequences with contrast medium, and natural sequences from the HEVC test data set~\cite{Bossen2013}. The dimensions are summarized in Table\,\ref{tab:data}. The bit depth for all sequences constitutes 8 bits per sample. 
All surveillance sequences are characterized by a static background and some moving objects in the foreground. The medical sequences origin from Digital Subtraction Angiography (DSA), showing the inflow of a contrast medium into a human cranium in frontal and sagittal perspectives. 

In the following, we will compare our proposed CA-WL to a uniform wavelet lifting (U-WL) with the same number of total decomposition levels. 
The single frames of each subband are encoded by JPEG\,2000, using the OpenJPEG~\cite{openjpeg} implementation with four spatial wavelet decomposition steps in $xy$-direction.
Further, we evaluate the CA-WL and the U-WL with and without a block-based MC, respectively. For block-based MC, the block size is set to 8, while the search range starts with a size of 8 and is doubled for every decomposition level until a maximum size of 64. The increasing search range is important, since the input frames of higher decomposition levels have a larger temporal distance, which has to be covered. To keep the computational effort realistic, we limit the increment of the search range by 64 pixels. The resulting motion vectors are encoded using the QccPack library~\cite{fowler2000qccpack}. 
Then, the entire file size is composed of the rate resulting from each subband, the required motion vectors and the coding costs for transmitting vector $\boldsymbol{v}$.
The visual quality of the resulting LP subband is measured by the same metric as already used in Section\,\ref{subsec:stoppingCrit}, but in terms of $\text{PSNR}_{\text{LP}_t}$~\cite{Lanz2018}. 

Table\,\ref{tab:results_all} gives the differences regarding $\text{PSNR}_{\text{LP}_t}$ in [dB] and the entire file size in [\%] of our proposed method compared to the U-WL for all data sets with and without the application of MC and for different values of $\lambda$. 
As can be seen in the right column, our method always results in a better visual quality compared to the U-WL. By increasing $\lambda$, the file size is reduced, while the $\text{PSNR}_{\text{LP}_t}$ gains are also decreased. However, for $\lambda{=}7$, the $\text{PSNR}_{\text{LP}_t}$ gains are still positive. 
By including MC into both methods, we are able to get a lower rate than for the U-WL, resulting in positive $\text{PSNR}_{\text{LP}_t}$ gains at the same time.
For $\lambda{=}3$, the file size can be reduced by up to $1.06\%$ in total average, while the visual quality is increased by $10.28$\,dB, as the right column of Table\,\ref{tab:results_all} shows.

To demonstrate the performance of our proposed CA-WL in more detail, Fig.\,\ref{fig:RD_results} presents the absolute rate and $\text{PSNR}_{\text{LP}_t}$ results from the \textit{AirportNight1} sequence with and without MC, using $\lambda{=}3$. As the left plot offers, the entire file size for incorporating MC is always higher than omitting MC. However, the visual quality is significantly higher by including MC, which is very important for many professional applications. But for higher decomposition levels $i$, the error propagation due to imperfect MC is increasing. The right plot shows the strong decreasing $\text{PSNR}_{\text{LP}_t}$ results, which can be prevented by our proposed CA-WL. Simultaneously, the overall file size can be decreased, as the left plot of Fig.\,\ref{fig:RD_results} shows. 

In Fig.\,\ref{fig:visual_results}, some visual results for each data set are presented. From left to right, the reference frame and the corresponding LP frames from the U-WL and the proposed CA-WL are shown for a value of $\lambda{=}3$, without MC and with block-based MC, respectively. 
Disturbing artifacts and loss of content, resulting from the U-WL, are highlighted in every frame. The rectangles depict blocking artifacts, the circles indicate locations of objects, which are canceled out completely, and the ellipses show blurring artifacts. As the right column shows, the CA-WL is capable to compensate this lack of data fidelity efficiently and gives a reliable impression of the actual content of the sequence. This is also proven by the PSNR values given at the bottom right corner of each frame.

\section{Conclusion}
\label{sec:conclusion}

Scalable lossless video coding and a high visual quality of the corresponding BL is very important for many professional applications. Wavelet-based video coding provides full scalability without additional overhead. The temporal resolution can be controlled by the recursive application of the WT in temporal direction to the LP subband of the previous stage. 
This leads to lower bit rates, but if the content of the underlying video sequence comprises strong motion, the visual quality of the BL is degraded significantly. 
We proposed a method which locally adapts the temporal scaling by evaluating a Lagrangian cost functional in every transformation step and prevents further decomposition, if the costs of the current level are higher than the costs of the previous level. This way, we can increase the visual quality of the BL by $10.28$\,dB compared to the U-WL with block-based MC, while the required rate is reduced by $1.06\%$ at the same time. Further work aims at the development of an algorithm to determine the optimum value of $\lambda$ in a rate-distortion sense, based on the characteristics of the underlying sequence.

\section{Acknowledgment}

We gratefully acknowledge that this work has been supported by the Deutsche Forschungsgemeinschaft (DFG) under contract number KA 926/4-3.

\vfill\pagebreak



\bibliographystyle{IEEEbib}
\bibliography{Literatur.bib}

\end{document}